\documentclass[showpacs, aps, prb, unsortedaddress]{revtex4-1}

\usepackage{amssymb}
\usepackage{latexsym}
\usepackage{amsmath}
\usepackage{graphicx}
\begin{document}

\title{Negative Refraction, Beam Steering, Mode Switching, and High-pass Filtering in a 1-D Periodic Laminate}

\date{\today}
\author{Ankit Srivastava}
\thanks{Corresponding Author}
\email{asriva13@iit.edu}

\affiliation{Department of Mechanical, Materials, and Aerospace Engineering
Illinois Institute of Technology, Chicago, IL, 60616
USA}

\begin{abstract}
In this paper we show that a 1-D phononic crystal (laminate) can exhibit metamaterial wave phenomenon which is traditionally associated with 2-, and 3-D crystals. Moreover, due to the absence of a length scale in 2 of its dimensions, it can outperform higher dimensional crystals on some measures. This includes allowing only negative refraction over large frequency ranges and serving as a near-omnidirectional high-pass filter up to a large frequency value. First we provide a theoretical discussion on the salient characteristics of the dispersion relation of a laminate and formulate the solution of an interface problem by the application of the normal mode decomposition technique. We present a methodology with which to induce a pure negative refraction in the laminate. As a corollary to our approach of negative refraction, we show how the laminate can be used to steer beams over large angles for small changes in the incident angles (beam steering). Furthermore, we clarify how the transmitted modes in the laminate can be switched on and off by varying the angle of the incident wave by a small amount. Finally, we show that the laminate can be used as a remarkably efficient high-pass frequency filter. An appropriately designed laminate will reflect all plane waves from quasi-static to a large frequency, incident at it from all angles except for a small set of near-normal incidences. This will be true even if the homogeneous medium is impedance matched with the laminate. Due to the similarities between SH waves and EM waves it is expected that some or all of these results may also apply to EM waves in a layered periodic dielectric.
\end{abstract}

\maketitle

\section{Introduction}

The observation that periodic structures can significantly affect wave propagation has inspired a large body of scientific research in the areas of both photonics and phononics over the last two decades. The excitement stems from the applications which have been proposed of these structures. Some of these include negative refraction, beam splitting, beam steering, and frequency filtering. Traditionally, the focus has been on 2-D and 3-D periodic structures with 1-D structures (laminates) serving merely as a stepping stone to more interesting problems in higher dimensions. What the laminates gained in the ease of their assembly, they lost in the apparent lack of richness in their wave physics. 

The body of literature is large and here we direct the reader to some review articles on these topics \cite{pennec2010two,lee2012micro,hussein2014dynamics,srivastava2015elastic}. This paper considers a specific problem within this field: oblique incidence of SH waves at an interface between a homogeneous material and a layered periodic composite. As such, the problem is similar to the propagation of EM waves in layered periodic dielectric. We assume that the layer interfaces are in the $x_2-x_3$ plane. The analysis constitutes determining the dispersion relation of the laminate for a wave-vector in the $x_1-x_2$ plane. Although a host of numerical techniques could be used, this problem is normally simple enough to admit an exact solution. The approach is well established in the form of the transfer matrix method (TMM). TMM can tackle both normal incidence \cite{lin1969dynamics,faulkner1985free,kutsenko2015tunable} and oblique incidence \cite{lekner1994light} with the former being a special case of the latter. In this paper we use the approach used by Lekner \cite{lekner1994light} for EM waves which was subsequently adapted by Willis recently for SH waves \cite{willis2015negative} in solids. The dispersion relation of the laminate in the context of mechanics has been studied by several authors over the last five decades \cite{sun1968time,nemat1972harmonic,nemat1975harmonic,hegemier1973continuum1,hegemier1973continuum2}. For the EM case the contributions go back even further with some of the first studies from Lord Rayleigh now more than a hundred years old \cite{rayleigh1887xvii,rayleigh1917reflection} (See also \cite{joannopoulos2011photonic}). That laminates exhibit dispersion and bandgaps has, therefore, been common knowledge for a considerable time now.

With the advent of photonic/phononic crystals and metamaterials, interest subsequently increased in the manifestation of exotic wave phenomenon including, but not limited to, negative refraction. The implicit assumption here is the existence of an interface of these crystals with another medium. The interface problem with a homogeneous material can be comprehensively studied through a combination of the TMM and normal mode decomposition. TMM provides information about not just the propagating modes but also the evanescent modes. Normal mode decomposition, on the other hand, tells us exactly how the energy is partitioned at the interface among the various available modes. With this, the study of such anti-plane shear waves (SH in solids or EM waves) can be broadly divided into two categories on the basis of the location of the interface between the homogeneous material and the laminate: a. when the interface lies in the $x_2-x_3$ plane and, b. when the interface lies in the $x_1-x_3$ plane. 

Of the two problem sets the former is the more conventional one \cite{boudouti2013one}. Within this, perhaps the most important (and most relevant here) observation, beyond the simple bandstructure effect, is the realization that laminates in this configuration can be used as omnidirectional reflectors \cite{winn1998omnidirectional,fink1998dielectric,bousfia2001omnidirectional,manzanares2004experimental}. The idea is to choose a combination of the frequency range, homogeneous material and the laminate such that the 2-component of the wave vector in the homogeneous material can never couple to a propagating mode in the laminate. In this paper we make an observation which, to the author's knowledge, has not been made in this interface configuration. The observation is that if we are ready to sacrifice the omnidirectionality of the reflector by a small amount, then we can instead create a highpass frequency filter using the laminate. Such a filter will reflect all waves from a quasi-static frequency to an arbitrarily large frequency value and for all angles except for an arbitrarily small range near normal incidence. The technique to do so is qualitatively different from designing for omnidirectionality as a different zone in the band-structure is utilized to achieve the effect.

The latter set of problems has received attention only recently through papers by Willis \cite{willis2015negative} and Nemat-Nasser \cite{nemat2015refraction,nemat2015anti}. It has been suggested that in this configuration it is possible to achieve negative refraction in a laminate - a phenomenon traditionally associated with more complex 2-, and 3-D crystals. Willis presented TMM+normal mode decomposition calculations and showed that at a chosen frequency the transmitted signal consisted of both positively and negatively refracted signals (beam splitting). In this paper we complement his and Nemat-Nasser's observations with several new observations which were not made in earlier papers. First we show that it is possible to achieve pure negative refraction in the laminate (no beam splitting with a positive refraction). This is achieved by coupling the wave in the homogeneous medium with the laminate modes \emph{not in the first} but in the second Brillouin zone at frequencies where only one propagating mode exists. This is clearly not possible when the interface is in the $x_2-x_3$ plane. Negative refraction achieved in this fashion is persistent over large ranges of angles of incidence and frequency. Second, we show that there exists frequencies where the positive and negative refraction angles are very large and that the beam can be made to traverse these large angles with only a small change in the incidence angle (beam steering). Finally, we show that the transmitted beams can be switched on and off through small changes in the angles of incidence.

In the following sections we present the relevant equations following Lekner \cite{lekner1994light} and Willis \cite{willis2015negative} (Sec. II). We complement these equations with theoretical discussions which illuminate certain essential characteristics of the dispersion relation. Poynting vector calculations are presented in Sec. III as the unabmiguous method of determining refractive possibilities of various modes. This treatment is different but equivalent to the EFC based arguments \cite{notomi2000theory}. In Sec. IV we summarize our observations for the case when the interface is in the $x_1-x_3$ plane (negative refraction, beam steering, and mode switching). TMM+normal mode decomposition equations are presented and energy based checks on the validity of the calculations are formulated. In Sec. IV we summarize our observation for the case when the interface is in the $x_2-x_3$ plane (high-pass filtering). 

\section{Relevant Equations}

Following \cite{willis2015negative} we define our laminate as a periodically layered structure in the $x_1$ direction with the layer interfaces in the $x_2-x_3$ plane and infinite in this plane. In the direction of periodicity the laminated composite is characterized by a unit cell $\Omega$ of length $h$ ($0\leq x_1\leq h$). For our purposes the unit cell is composed of 2 material layers with shear moduli $\mu_1,\mu_2$, density $\rho_1,\rho_2$, and thicknesses $h_1,h_2$ respectively. The case of $n$ homogeneous material layers or layers with spatially changing material properties is not substantively different. All that is required is that the material properties be periodic with the unit cell so as to give the composite its phononic character.

If anti-plane shear waves are propagating in the laminate then the only nonzero component of displacement is taken to be $u_3$ which has the functional form $u_3(x_1,x_2,t)$. Within the $i^{th}$ layer ($i=1,2$) it satisfies the following equation of motion:
\begin{eqnarray}
\label{eAntiPlane}
\displaystyle u_{3,11}+u_{3,22}=\frac{1}{c^2_i}\ddot{u}_3
\end{eqnarray}
where $c_i=\sqrt{\mu_i/\rho_i}$. The displacement gives rise to stress fields $\sigma_{13}(x_1,x_2,t),\sigma_{23}(x_1,x_2,t)$. The shear stress component $\sigma_{13}$ and displacement $u_3$ are continuous at the material interfaces. Across an interface between layers $i$ and $i+1$ at $x_1=x^i$:
\begin{equation}
\label{eContinuity}
\mathbf{v}^i|_{x_1=x_i}\equiv\begin{pmatrix}\sigma_{13}(x^i,x_2,t) \\ u_3(x^i,x_2,t) \end{pmatrix}^i=\mathbf{v}^{i+1}|_{x_1=x_i}
\end{equation}
Due to the periodicity of the laminate, the displacement and stress fields follow Bloch-periodicity conditions. Generally we have $\mathbf{v}\equiv\tilde{\mathbf{v}}(x_1)e^{i(\omega t-K_1x_1-k_2x_2)}$ and, for the displacement field, we have:
\begin{eqnarray}
\label{eBlochAntiPlane}
\displaystyle u_3(x_1,x_2,t)=\tilde{u}(x_1)\mathrm{e}^{i(\omega t-K_1x_1-k_2x_2)}
\end{eqnarray}
where $\tilde{\mathbf{v}}(x_1)$ is periodic with $\Omega$. The wavenumber component $k_2$ must be continuous across the layers to satisfy Snell's law. Similar Bloch periodicity relations apply to the other nonzero stress component $\sigma_{23}$ as well. By using the general solutions to the governing equation (\ref{eAntiPlane}), the continuity of stress and displacement at the interfaces (\ref{eContinuity}), and the Bloch formulation (\ref{eBlochAntiPlane}), we can formulate a Transfer Matrix formulation ($x_2,\omega$ dependence suppressed):
\begin{equation}
\label{eTMM}
\mathbf{v}(h)=M\mathbf{v}(0)=\lambda\mathbf{v}(0)
\end{equation}
where the eigenvalue $\lambda=e^{-iK_1h}$. Quantities in the above equation depend upon assumed values of $\omega,k_2$. The solutions to the eigenvalue problem above furnish the wavenumber $K_1$ and the modeshape for which (\ref{eBlochAntiPlane}) satisfies the governing equation. The wavenumber solutions themselves come from the following equation:
\begin{equation}
\label{eAntiPlaneS}
\cos(K_1h)=\frac{1}{2}\mathrm{tr}(M)
\end{equation}
so that if $K_1$ is a solution then so are $\pm(K_1\pm 2n\pi/h)$ for all integer $n$.

\subsection{Solution Spectrum}
The dispersion relation consists of the solutions of Eq. (\ref{eAntiPlaneS}) for assumed values of $\omega$ and $k_2$. The equation itself is valid for real, imaginary, or complex values of frequency and wavenumbers. For real and positive values of frequency the solution spectrum can be divided on the basis of whether $k_2$ is real (propagating wave) or imaginary (evanescent wave). For real and positive values of $k_2$ (and an assumed value of frequency) it can be shown that there exists an upper limit of $k_2$ beyond which the associated $K_1$ solutions are purely imaginary. From Eq. (\ref{eAntiPlaneS}), $K_1$ will be purely imaginary when the trace of $M$ is real and greater than 2. Following \cite{lekner1994light} and after some algebraic manipulations we can write down this condition:
\begin{eqnarray}
\label{eImKC}
\displaystyle 2\cos(\phi_1)\cos(\phi_2)-\sin(\phi_1)\sin(\phi_2)\left[f+\frac{1}{f}\right]>2;\quad f=\frac{\mu_1q_1}{\mu_2q_2};\quad q_i=\sqrt{\frac{\omega^2}{c_i^2}-k_2^2};\quad \phi_i=q_ih_i
\end{eqnarray}
Clearly, for a large enough $k_2$ both $q_1,q_2$ are purely imaginary and, if positive roots are taken for the sake of this argument, then $f$ is positive and real. Moreover for large $k_2$, $f\approx \mu_1/\mu_2$ and $f+1/f$ is at least as big as 2. We now express $f+1/f$ as $2+\alpha$ where $\alpha$ is a positive and real number and write the L.H.S of the above equation as:
\begin{eqnarray}
\displaystyle 2\cos(\phi_1+\phi_2)-\alpha\sin(\phi_1)\sin(\phi_2)
\label{eImKC1}
\end{eqnarray}
Since both $\phi_1,\phi_2$ are purely imaginary $2\cos(\phi_1+\phi_2)$ is bounded from below by a value of 2. Moreover, since $\alpha$ is positive and $\phi_1,\phi_2$ are positive and imaginary, $\alpha\sin(\phi_1)\sin(\phi_2)$ is negative and real. Therefore, L.H.S in Eq. (\ref{eImKC}) is greater than the R.H.S. The argument holds identically if negative imaginary roots are chosen for $q_1,q_2$ since $f$, in the limit, is still $\mu_1/\mu_2$. If a positive imaginary root and a negative imaginary root are chosen then $f$ approaches $-\mu_1/\mu_2$ and the left hand side approaches $2\cos(\phi_1-\phi_2)+\alpha\sin(\phi_1)\sin(\phi_2)$ where $\alpha$ is again positive and real. The cosine term is again greater than 2 and since one of $\phi_1,\phi_2$ is positive imaginary and the other is negative imaginary $\alpha\sin(\phi_1)\sin(\phi_2)$ is again greater than 0. Therefore, for all possible cases, the left hand side of the inequality in Eq. (\ref{eImKC}) is greater than the right hand side for large enough values of $k_2$. Eq. (\ref{eImKC}) is, therefore, identically satisfied for large enough values of $k_2$ or, in other words, $K_1$ is purely imaginary beyond a certain maximum value of $k_2$. This maximum value depends upon the frequency under consideration and increases with the frequency. Another interesting corollary of this analysis is when $\omega/c_i\ll k_2$. Again in this case both $\phi_1,\phi_2$ are imaginary and $f$ is either $\mu_1/\mu_2$ or $-\mu_1/\mu_2$. Following the above rationale we can say that even in this case the only possible $K_1$ solution will be an imaginary one. In other words, for non-zero values of $k_2$, there exists a special stopband for waves traveling in the $x_1$ direction. This stopband begins at 0 frequency and stretches up to some maximum frequency value. Such a quasi-static stopband is impossible to generate when waves normal to the layers are considered and its existence was noted by Willis \cite{willis2015negative} for their example. Here we have shown that this must be the case for all 2-material layered composites.
\begin{figure}[htp]
\centering
\includegraphics[scale=.65]{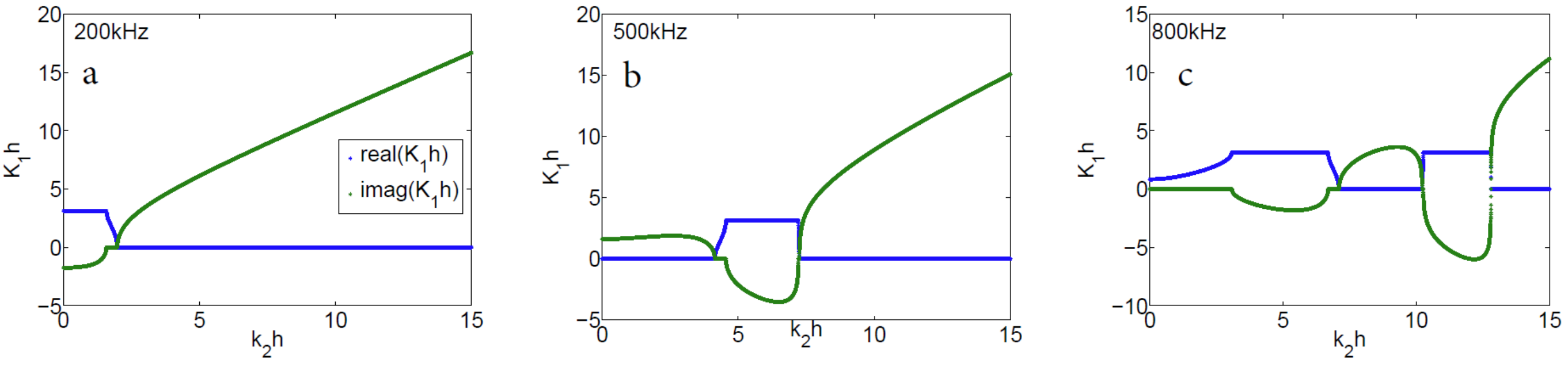}
\caption{$K_1h-k_2h$ plot for the laminate at three different frequencies. All $k_2h$ solutions are real.}\label{fk2M}
\end{figure}

As an example we consider a periodically layered composite made up of two materials with properties $\mu_1=80$ GPa, $\rho_1=8000$ kgm$^{-3}$, $h_1=.0013$ m and $\mu_2$=3 GPa, $\rho_2$=1180 kgm$^{-3}$, $h_2=.003$m. This composite has been considered previously in other papers as well \cite{nemat2011negative,srivastava2014limit}. Fig. (\ref{fk2M}) shows the $K_1-k_2$ solutions for real and positive values of $k_2$ and for three different freqeuencies. $K_1h$ has been restricted to the first Brillouin zone ($\mathcal{R}(K_1h)<\pi$) with the understanding that if $K_1$ is a solution then so are $\pm(K_1\pm 2n\pi/h)$ for all integer $n$. It is clear from the figure that as $k_2$ increases beyond a certain maximum $K_1$ becomes imaginary and remains imaginary. The wave is evanescent in the $x_1$ direction at this point. Below this maximum value there can be several real $k_2$ solutions for a given real $K_1$ (and also several real $k_2$ solutions for a given imaginary $K_1$). This follows from the additional observation that there are no complex $K_1$ solutions. The wave is either propagating or evanescent. It is never propagating and dissipative as there is no mechanism in the system to account for any dissipation. The number of such real $k_2$ solutions again depends upon the frequency under consideration and this number increases with the frequency. For instance, in the 200 kHz case there is only one real $k_2$ solution for a given real $K_1$ value, whereas, in the 800 kHz case the number of such solutions increases to 4. As we consider higher and higher frequencies we should expect to get more propagating wave modes in both $x_1,x_2$ directions. 
\begin{figure}[htp]
\centering
\includegraphics[scale=.4]{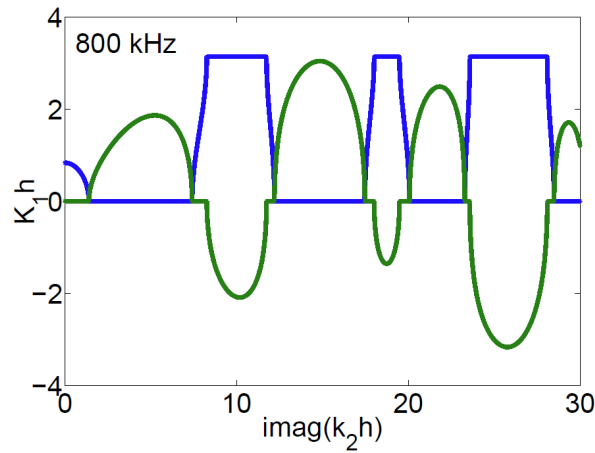}
\caption{$K_1h-k_2h$ plot for the laminate at 800 kHz. All $k_2h$ solutions are imaginary.}\label{800kHzIm}
\end{figure}

In summary, for a real $K_1$ value and at a frequency $\omega$ we expect $n_\omega^{rr}$ real $k_2$ solutions where $n_\omega^{rr}$ is a finite whole number. These constitute all allowable propagating waves in both $x_1$ and $x_2$ directions. In addition, for an imaginary $K_1$ value and at frequency $\omega$ we expect $n_\omega^{ir}$ real $k_2$ solutions where $n_\omega^{ir}$ is a finite whole number. These constitute the allowable waves which propagate in the $x_2$ direction but which are evaescent in the $x_1$ direction. Contrary to these, given a real (or imaginary) value of $K_1$, we have an infinite number of imaginary $k_2$ solutions. The reason that $n_\omega^{rr},n_\omega^{ir}$ are finite numbers is because for real values of $k_2$, the trace of $M$ is a monotonically increasing function beyond some maximum value of $k_2$. This is evident from the form of Eq. (\ref{eImKC1}). When $k_2$ is imaginary then $q_i,\phi_i$ are real quantities and the trace of $M$ is always bounded and oscillatory with $k_2$. This results in the existence of an infinte number of real $K_1$, imaginary $k_2$ and imaginary $K_1$, imaginary $k_2$ solution sets. The former set constitutes waves which are propagating in the $x_1$ direction and evanescent in the $x_2$ direction and the latter set constitutes waves which are evanescent in both directions. Fig. (\ref{800kHzIm}) shows the calculated solutions for imaginary $k_2$ (up to $k_2h=30i$) for the composite under consideration. More solutions should be expected as one considers larger values of imaginary $k_2$.

\section{Poynting Vector}
We also want to determine the direction of time and unit cell averaged energy flux which is related to the Poynting vector. The time averaged real part of the Poynting vector gives the energy flux:
\begin{equation}
\boldsymbol{\mathcal{P}}=-\frac{1}{2}\mathcal{R}\left[\boldsymbol{\sigma}\cdot\dot{\mathbf{u}}^*\right]
\end{equation}
which in the present case has two non-zero components:
\begin{eqnarray}
\displaystyle \mathcal{P}_1=\frac{1}{2}\mathcal{R}\left[\omega\mu(K_1|\tilde{u}|^2+i\tilde{u}'\tilde{u}^*)\right];\quad \mathcal{P}_2=\frac{1}{2}\mathcal{R}\left[\omega \mu k_2 |\tilde{u}|^2\right]
\end{eqnarray}
The unit cell averages of the above (denoted by the brackets $\langle \rangle$) give the time and unit cell averaged energy flux in the composite and indicate the direction of energy flow. Clearly, $\langle \mathcal{P}_2\rangle$ will always have the sign of the real part of $k_2$. $\langle \mathcal{P}_2\rangle$ will be zero if $k_2$ is purely imaginary as in the case of evanescent waves in the $x_2$ direction. Moreover, the sign of $\mathcal{P}_2$ is always the same as $\langle \mathcal{P}_2\rangle$ indicating that the energy transport in the $x_2$ direction at the micro scale is always in the same direction as at the macro scale. The case is more complicated for $\mathcal{P}_1$ whose sign is determined, both at the micro and macro scales, by the relative magnitude of $\mu\mathcal{R}[i\tilde{u}'\tilde{u}^*]$ with respect to $\mu\mathcal{R}[K_1|\tilde{u}|^2]$.
\begin{figure}[htp]
\centering
\includegraphics[scale=.4]{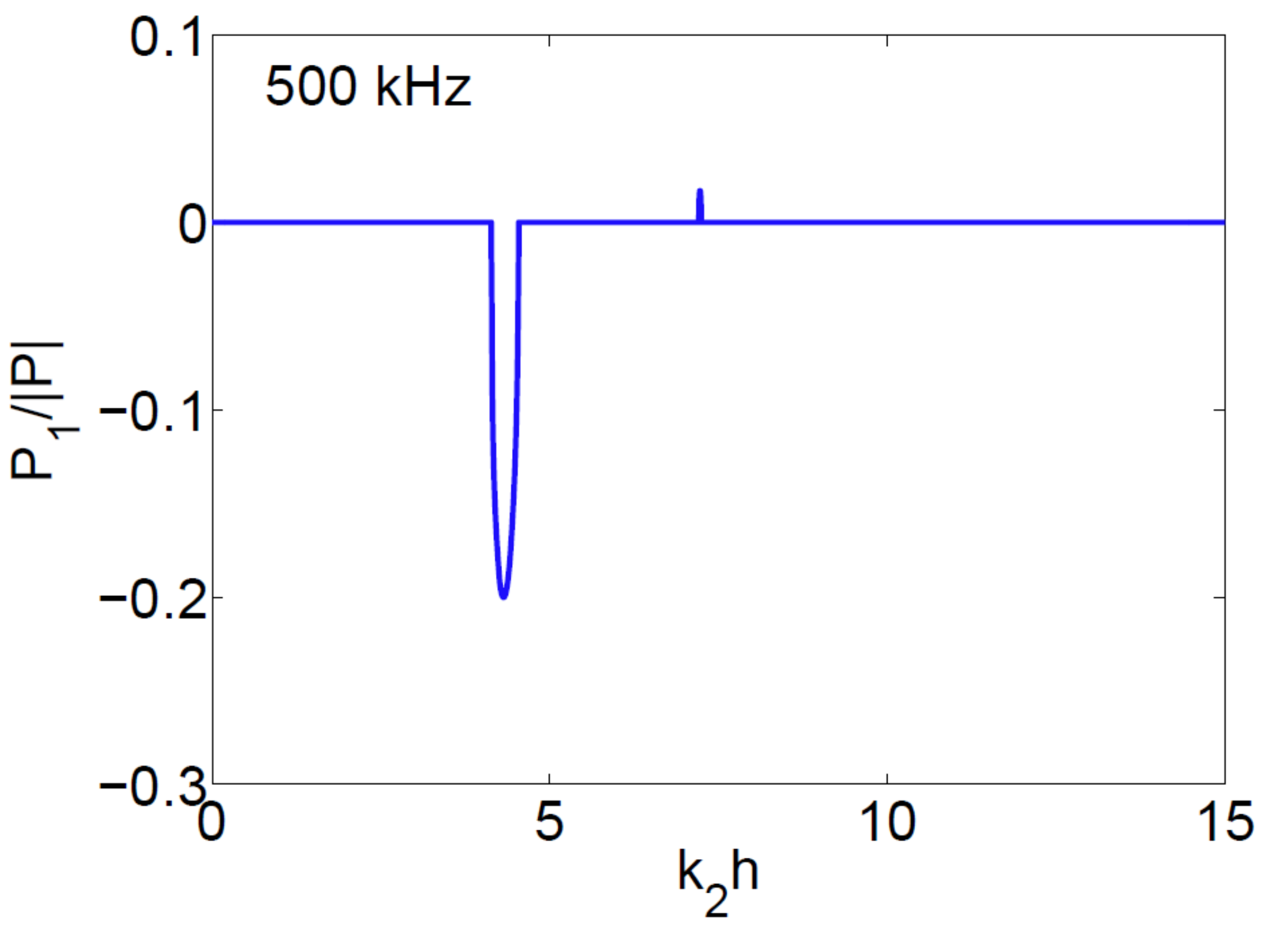}
\caption{The 1-component of the Poynting vector as a function of real $k_2h$ at 500 kHz.}\label{Poynting}
\end{figure}

Fig. (\ref{Poynting}) shows the calculated value of $\mathcal{P}_1$ for real $k_2$ solutions at the frequency of 500 kHz. When compared with the 500 kHz case in Fig. (\ref{fk2M}) it is clear that $\mathcal{P}_1$ is 0 when $\mathcal{R}(K_1)$ is either 0 or $\pi$. For the first passband (lower $k_2h$ solution) it is negative and for the second passband it is positive. The second passband is the zone of traditional refraction and the first passband corresponds to the zone where negative refraction is possible. Similarly Poynting vector calculations for imaginary $k_2$ case also reveal bands with positive ($\mathcal{R}(K_1)\mathcal{P}_1>0$) and negative refraction ($\mathcal{R}(K_1)\mathcal{P}_1<0$) properties. However, since $k_2$ is imaginary for this case, these waves do not carry energy in the $x_2$ direction ($\mathcal{P}_2=0$). They will be generated at appropriate interfaces and will travel along those interfaces, as will be shown later. It can be shown that the sign of $\langle\mathcal{P}\rangle$ is the same as the sign of $\partial\omega/\partial \mathcal{R}(K_1)$ \cite{willis2015negative}. Given the nature of the dispersion curves, this indicates that if $\langle\mathcal{P}\rangle$ is positive (negative) for some $K_1$ then it will be positive (negative) for all $K_1+2n\pi$ solutions and negative (positive) for all $-K_1+2n\pi$ solutions. 

\section{Negative Refraction, Beam Steering, and Mode Switching}

Here we show that not only is negative refraction possible in the simple 1-D phononic crystal (as originally demonstrated by Willis) but under appropriate conditions a wave can only refract negatively in the laminate. Moreover, this negative refraction is remarkably robust and persistent over large frequency ranges and angles of incidence.

First consider two different solutions of the current problem (details in \cite{willis2015negative}):
\begin{eqnarray}
\displaystyle u_3(x_1,x_2,t)=\tilde{u}(x_1)\mathrm{e}^{i(\omega t-K_1x_1-k_2x_2)}\\
\displaystyle v_3(x_1,x_2,t)=\tilde{v}(x_1)\mathrm{e}^{i(\omega t-K_1x_1-\bar{k}_2x_2)}
\end{eqnarray}
It can be shown that as long as $\omega^2,K_1,k_2^2,\bar{k}_2^2$ are real and $k_2^2\neq \bar{k}_2^2$, the modeshapes $\tilde{u},\tilde{v}$ are orthogonal with respect to the weight $\mu$:
\begin{eqnarray}
\label{ortho}
\int_0^h\tilde{u}\mu\tilde{v}^*\mathrm{d}x_1=0
\end{eqnarray}
The above requirements are satisfied by all the solutions which are considered in the previous sections. 
\begin{figure}[htp]
\centering
\includegraphics[scale=.8]{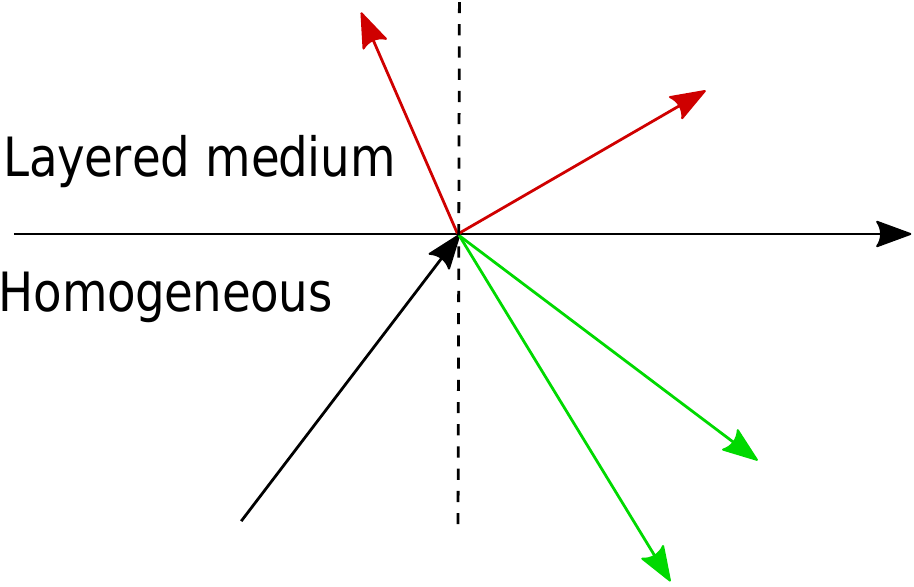}
\caption{Schematic of the interface problem. The interface is along $x_2=0$.}\label{layered}
\end{figure}
We now consider an interface between a homogeneous medium with shear modulus $\mu_0$ and the layered composite. The interface itself can be placed at any angle with the layers but presently we assume that it is along $x_2=0$ (Fig. \ref{layered}). The layered medium is in the region $x_2>0$ with the layers being parallel to the $x_2$ axis. This is the configuration which was studied in \cite{willis2015negative} where it was noted that it allows for the possibility of negative refraction. A plane harmonic wave is incident at the interface from the homogeneous medium. This wave sets up an infinite number of transmitted and an infinite number of reflected waves. A finite number of these are propagating waves and the rest are evanescent waves. Fig. (\ref{layered}) shows 2 propagating transmitted and reflected waves but in general there can be more such propagating scatterings. The incident, transmitted, and reflected fields are written down as:
\begin{eqnarray}
\displaystyle \mathrm{Incident Field:}\quad A\exp\left[i(\omega t-k\sin\theta x_1-k\cos\theta x_2)\right]\\
\displaystyle \mathrm{Transmitted Field:}\quad \sum_{m=0}^\infty T_m\tilde{u}_m(x_1)\exp\left[i(\omega t-k\sin\theta x_1-k_m x_2)\right]\\
\displaystyle \mathrm{Reflected Field:}\quad \sum_{n=-\infty}^\infty R_nU_n(x_1)\exp\left[i(\omega t-k\sin\theta x_1+\kappa_{n} x_2)\right]
\end{eqnarray}
where $U_n(x_1)=e^{-i2n\pi x_1/h}$ and $\kappa_{n}=\left[k^2-(k\sin(\theta)+2n\pi/h)^2\right]^{1/2}$ and $\kappa_{n}$ is taken either as positive real or negative imaginary to prevent exponential rise in $x_2<0$. At any given frequency, the transmitted field will consist of $M$ propagating modes and infinitely many evanescent modes in the $x_2$ direction. To facilitate calculations we restrict the reflected modes to a range of $-N\leq n\leq N$ and transmitted modes to a range of $0\leq m\leq 2N$ such that $2N+1>M$. This allows us to consider all propagating transmitted modes in the expansion. With this, the displacement ($u_3$) and stress ($\sigma_{32}$) continuity are given by (exponential terms suppressed):
\begin{eqnarray}
\displaystyle \sum_{m=0}^{2N}\bar{T}_m\tilde{u}_m(x_1)-\sum_{n=-N}^{N}\bar{R}_nU_n(x_1)=1\\
\displaystyle \mu(x_1)\sum_{m=0}^{2N}k_m\bar{T}_m\tilde{u}_m(x_1)+\mu_0\sum_{n=-N}^{N}\kappa_{n}\bar{R}_nU_n(x_1)=\mu_0k\cos\theta
\end{eqnarray}
where $\bar{R}_n=R_n/A,\bar{T}_n=T_n/A$. The above can be transformed into a system of $2(2N+1)$ equations in as many variables through the application of the orthogonality condition (\ref{ortho}). Specifically we have
\begin{eqnarray}
\label{scattering}
\displaystyle 
\begin{bmatrix}
    [M_1] & [M_2]\\
    [M_3] & [M_4]
\end{bmatrix}\{S\}=\{I\}
\end{eqnarray}
where $S$ is a column vector of size $2(2N+1)$ with elements $\bar{T}_0,...\bar{T}_{2N},\bar{R}_{-N},...\bar{R}_{N}$. Submatrices $[M_i]$ are square matrices of sizes $(2N+1)\times(2N+1)$ with the following nonzero elements:
\begin{eqnarray}
\displaystyle [M_1]_{ii}=\int_0^{h}\mu\tilde{u}_i\tilde{u}_i^*\mathrm{d}x_1,\quad [M_2]_{ij}=-\int_0^{h}\mu U_{j-N}\tilde{u}_i^*\mathrm{d}x_1\\
\displaystyle [M_3]_{ii}=k_i\int_0^{h}\mu \tilde{u}_i\tilde{u}_i^*\mathrm{d}x_1,\quad [M_4]_{ij}=\mu_0 \kappa_{j-N}\int_0^{h}U_{j-N}\tilde{u}_i^*\mathrm{d}x_1\\
i,j=0,...2N
\end{eqnarray}
and $I$ is a column vector of size $2(2N+1)$ with elements
\begin{eqnarray}
\displaystyle I_{i}=\int_0^{h}\mu\tilde{u}_i^*\mathrm{d}x_1,\quad 0\leq i\leq 2N\\
\displaystyle =\mu_0k\cos(\theta)\int_0^{h}\tilde{u}_i^*\mathrm{d}x_1,\quad i>2N
\end{eqnarray}

\subsection{Energy Considerations}
As a check on the consistency of the calculations we also need to consider the balance of energy flow in the system. Fig. (\ref{layered}) shows the schematic of the problem under consideration. The green arrows represent the energy flow direction of the propagating reflected waves and the red arrows represent the energy flow directions of the propagating transmitted waves. These directions are generally not the same as the directions of the corresponding wave-vectors and are instead given by the directions of their Poynting vectors. Now consider a rectanglular area of length $h$ and height $t$ with its latter dimension bisected by $x_2=0$. In the absence of any dissipating mechanisms the total energy entering this rectangle should balance the energy leaving it. The average energy entering this area due to the incident wave is $A^2\mu_0\omega kh\cos\theta_i/2$ where $\theta_i$ is the angle of incidence. There is a net loss of energy from this region due to the presence of propagating transmitted and reflected waves. Energy contained in these waves is generated at the interface and then transported away. The waves which are evanescent in the $x_2$ direction do not affect the net balance of energy in this region as they have to both enter and exit it. Since the scattering coefficients are given in a combined fashion by the $\mathbf{S}$ vector in (\ref{scattering}), we can write down energy balance based on these coefficients. If the 2-component of the time and unit cell averaged Poynting vector for the $i^\mathrm{th}$ wave in $\mathbf{S}$ is given by $\langle\mathcal{P}\rangle_2^{(i)}$ then it represents a net time averaged energy loss of $|S_i|^2\langle\mathcal{P}\rangle_2^{(i)}h$ from the system. Energy balance equation thus becomes:
\begin{eqnarray}
\label{energyidentity}
E=\frac{2}{\mu_0\omega k\cos\theta_i}\sum_{i=1}^{2(2N+1)}|\bar{S}_i|^2\langle\mathcal{P}\rangle_2^{(i)}=1
\end{eqnarray}
It should be noted that the above equation will only have contributions from the propagating waves since the 2-component of the Poynting vector for any wave which is evanescent in $x_2$ will be zero. This does not mean, however, that only propagating waves are generated. The total energy flux for the incident wave is given by the magnitude of its time averaged Poynting vector ($A^2\mu_0\omega k/2$). The energy represented by this flux is distributed between the transmitted and reflected propagating and evanescent waves. The 2-component of the energy flux for an evanescent wave is obviously zero but the 1-component will be proportional to $\exp(-2k_2x_2)$ for $x_2>0$ and $\exp(2k_2x_2)$ for $x_2<0$ where $k_2>0$. This signifies that the evanescent waves will carry energy in the $x_1$ direction but that their flux will exponentially diminish away from $x_2=0$. In fact, if one of the evanescent reflected waves in (\ref{scattering}) has a coefficient $R$ then this wave represents a time averaged energy of $|R|^2\mu_0\omega k/(4k_2)$ crossing $x_1=0$. Eq. (\ref{energyidentity}) is used as the check on the validity of the calculations. For transmitted and reflected waves we would like to determine what fraction of the incident energy is converted into each scattered mode. This can be determined by calculating the normalized time and unit cell averaged values of the magnitude of the Poynting vector for each of the propagating scattered modes. The normalization will be done with respect to the magnitude of the Poynting vector of the incident wave.

\subsection{Examples}

First we consider the general example that was treated in \cite{willis2015negative}. In that reference the scattering parameters were calculated for one frequency and one angle of incidence and evanescent waves were not considered. The homogeneous medium is taken to be Aluminum ($\mu_0=26 \mathrm{GPa},\rho_0=2700 \mathrm{kg/m}^3$). We take the frequency of excitation to be 500 kHz in which case the first two propagating bands are fully developed (Fig. \ref{fk2M}b). The speed of the shear wave in Aluminum is 3103 m/s and at 500 kHz its wave number is $k=1012.4$/m. For the solution to be in the first Brillouin zone of the layered composite ($K_1h\leq \pi$) the angle of incidence is, therefore, limited to $\theta\leq 46.1922$ degrees. Within this zone there are two propagating modes in the layered composite. The mode with the smaller $k_2h$ ($T_0$ mode) refracts negatively and the other ($T_1$ mode) refracts positively (Fig. \ref{Poynting}). Additionally, there are infinite solutions which are evanescent in the $x_2$ direction. We take $N=7$ to calculate the scattering coefficients in Eq. (\ref{scattering}). This takes into account 2 propagating transmitted modes and 13 evanescent transmitted modes. The number of propagating reflected modes is determined by the value of the incident angle.
\begin{figure}[htp]
\centering
\includegraphics[scale=.5]{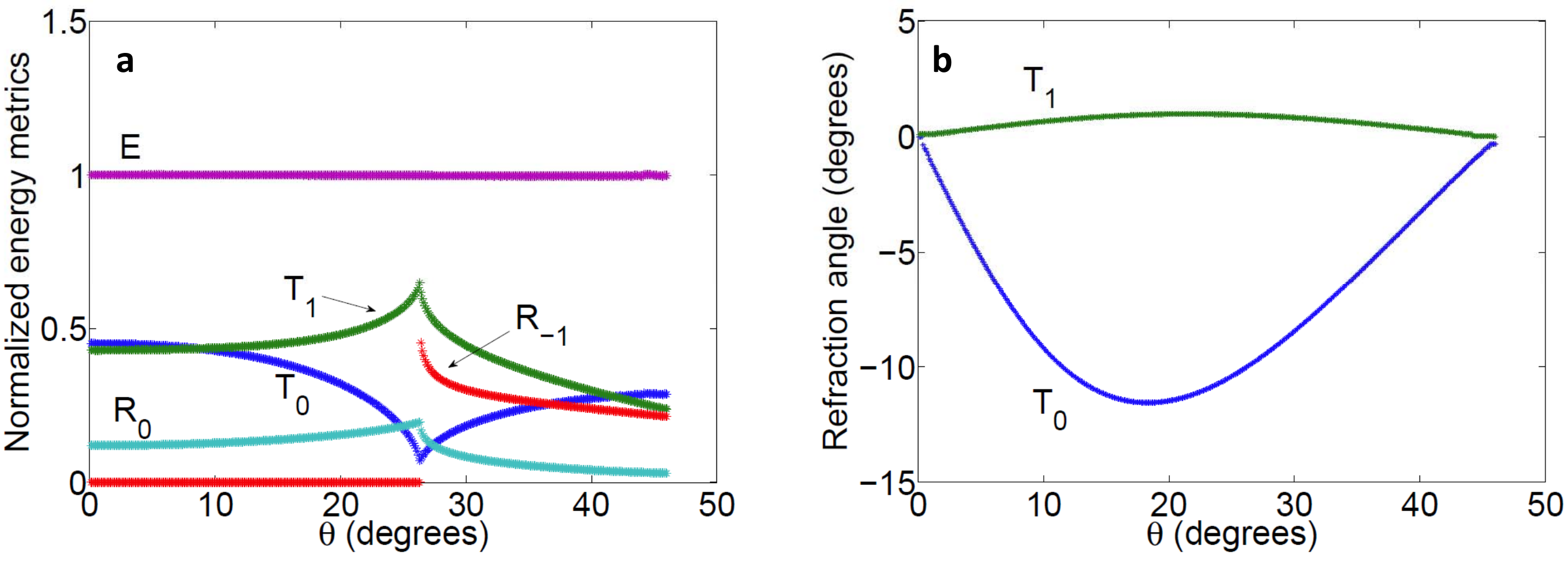}
\caption{a. Normalized energy metrics for the transmitted and reflected propagating modes, b. Angles of refraction for the energies of the transmitted modes.}\label{L45}
\end{figure}

Fig. (\ref{L45}) shows the normalized energy metrics for the problem for $0<\theta\leq 46^0$. $E$ refers to the energy metric from Eq. (\ref{energyidentity}) calculated for each incident angle. It can be seen that $E$ is very close to 1 for all cases, as expected. The other curves show the time and unit cell averaged energy flux magnitudes for the propagating transmitted and reflected modes normalized by the energy flux of the incident mode. These are referred to by their labels in the $\mathbf{S}$ vector, however, it is understood that their energy is meant. The $R_{-1}$ mode is evanescent before $\theta=26.4^0$, however, this critical angle will, in general, depend upon both the material properties of the homogeneous medium and $h$. There is a qualitative change in the behavior of the modes at the critical angle for the $R_{-1}$ mode. While the relative energies which reside in the $R_0$ and $T_1$ modes attain a maximum, the energy in the negatively refracting $T_0$ mode attains a minimum at this angle. For this entire incident angle range, the angles of refraction (energy) for the two transmitted modes are smaller than $12^0$ from the normal as shown in the adjacent figure.
\begin{figure}[htp]
\centering
\includegraphics[scale=.45]{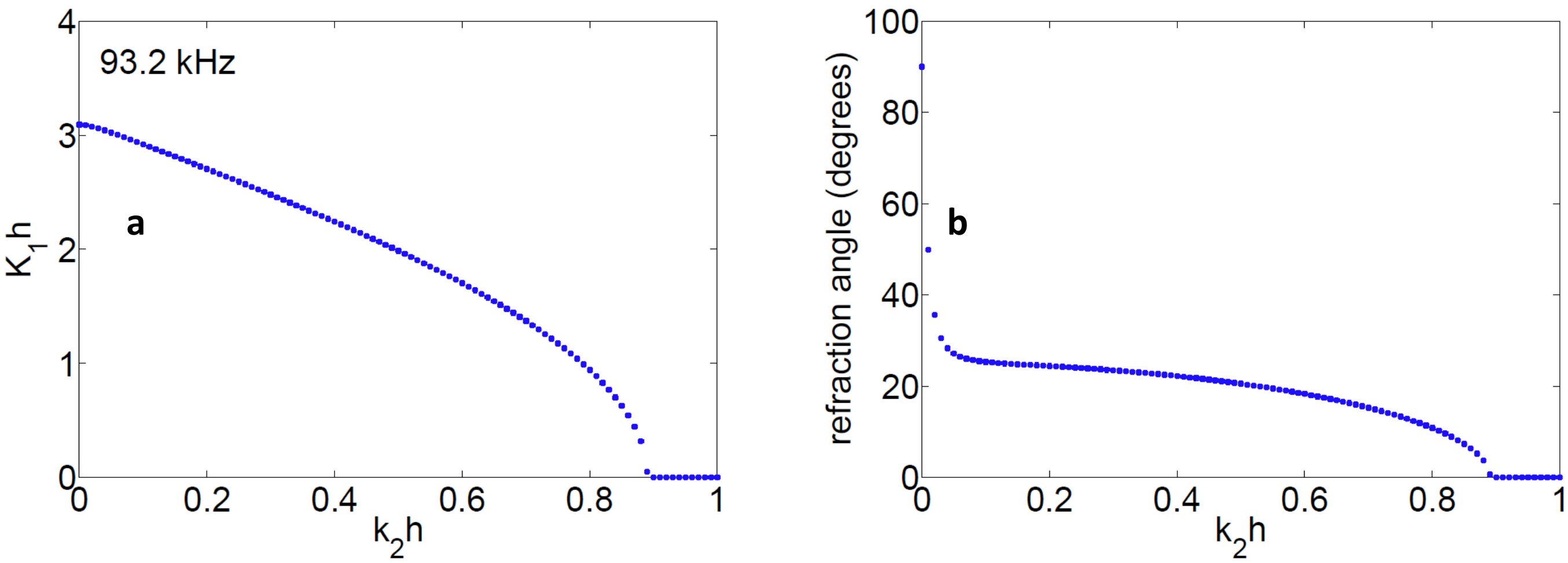}
\caption{a. $K_1h-k_2h$ plot at 93.2 kHz, b. Angle of refraction for the energy of the propagating mode.}\label{932kHz}
\end{figure}

It is clear from the above example that the refracted signal will, in general, be composed of both the positively and negatively refracting modes. It is possible to generate only one mode if the incident wave could appropriately couple to its modeshape, which may be possible through the use of transducer arrays. However, even if such a coupling could be established the refraction angles will still be small. Can we induce only a negative refraction in the layered composite with a large refraction angle? This is possible by considering a frequency where only one mode is developed and by simultaneously having an incident angle which couples to that mode \emph{not in the first} but in the second Brillioun zone. Moreover, the frequency should be such that the mode refracts at a large angle with the normal. Consider, for instance, the $K_1h-k_2h$ plot for the composite at 93.2kHz (Fig. \ref{932kHz}a).
\begin{figure}[htp]
\centering
\includegraphics[scale=.5]{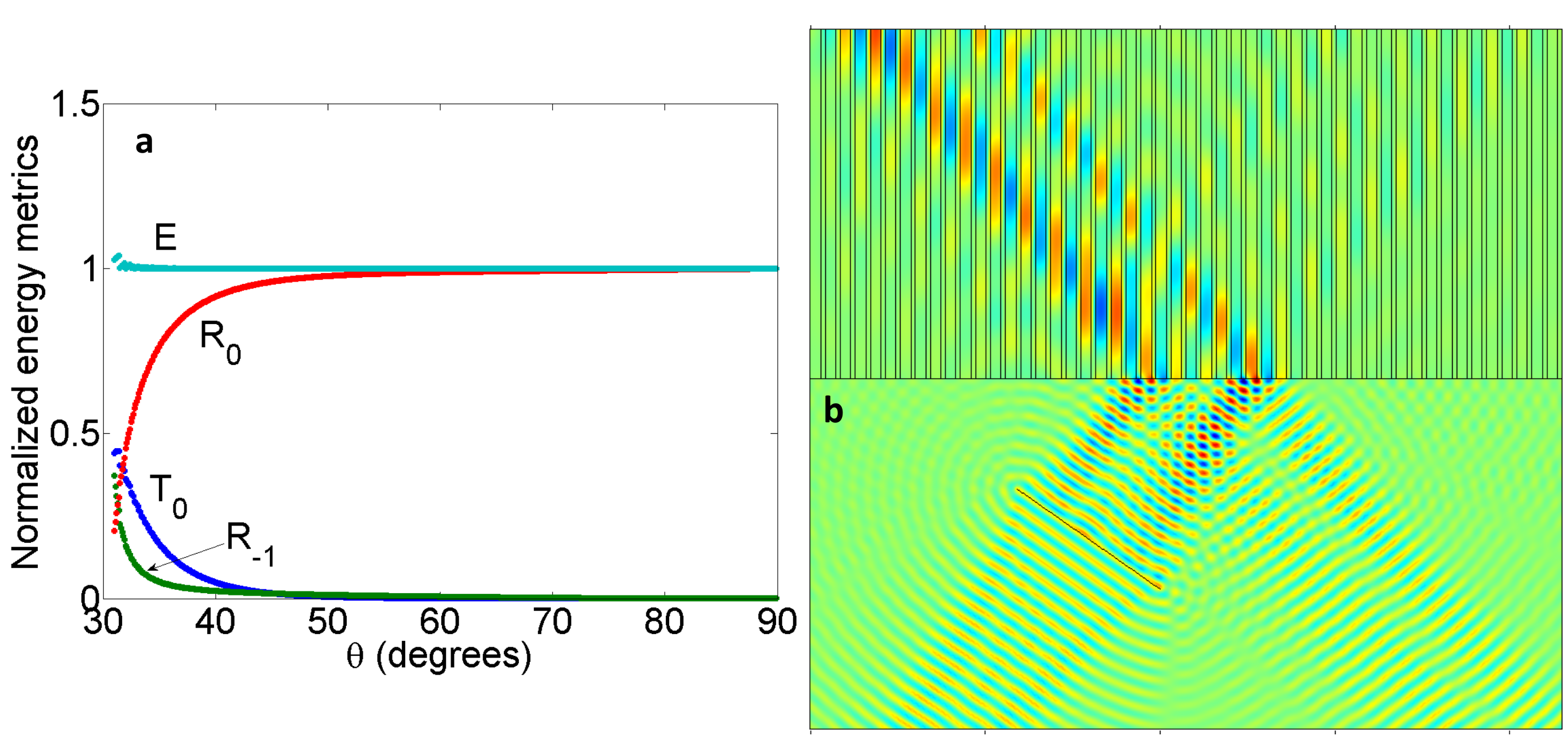}
\caption{a. Normalized energy metrics for the transmitted and reflected propagating modes at 93.2 kHz, b. Frequency domain Comsol simulation showing negative refraction in the laminate.}\label{G30SB}
\end{figure}

At this frequency only one mode is developed which refracts positively. However, due to the periodicity of the composite $-K_1h+2\pi$ are also solutions of the problem for a given $k_2$ and these solutions refract negatively by the same magnitude as the $K_1h$ solutions do. This means that for $\pi<kh\sin\theta<2\pi$ there will only be a negatively refracted propagating wave in the layered composite. The largest such range of $\theta$ will exist when $\pi/(kh)=0.5$, in which case $30^0<\theta<90^0$ will couple the incident wave to the mode in the second Brillouin zone. For smaller values of $\pi/(kh)$, negative refraction will begin at smaller incident angles and the range of incident angles over which negative refraction will be manifested will also be smaller. For this limiting case $\pi/(kh)=0.5$ determines the speed of the shear wave in the homogeneous medium. Specifically, at 93.2 kHz, the homogeneous material should be such that the speed of the wave is 400.76 m/s. Assuming a density of $3000$ kg/m$^3$, this results in the requirement of a material which is fairly compliant ($\mu_0=0.4818$ GPa). We note here that merely changing the scale of the problem by varying $h$ will have no effect on these requirements from the homogeneous material. For now we assume these as the properties of the homogeneous material. 

Fig. (\ref{G30SB}a) shows the transmission and reflection metrics for the current problem for an incident angle range of $30^0<\theta<90^0$. The only transmission is the $T_0$ mode which refracts negatively. Its energy decreases monotonically as $\theta$ increases. Fig. (\ref{G30SB}b) shows a frequency domain Comsol simulation of the problem under consideration. A total of 50 unit cells is used to model the layered composite and the homogeneous material has properties $\rho_0=3000$ kg/m$^3$ and $\mu_0=.4818$ GPa. A line load approximates a plane wave traveling towards the interface at an incident angle of $35^0$ with the normal. The domain boundaries have non-reflecting boundary conditions. As can be seen from the figure, at the simulation frequency of 93.2 kHz the predominant refraction is negative. Only a very small part of the energy refracts positively. However, that may very well be due to the wave not being exactly plane, with some of its energy impinging the interface at other angles. It should also be noted that the reflection is primarily the $R_0$ mode, which should be expected from Fig. (\ref{G30SB}a) at $35^0$ incidence. Pure negative refraction of the kind described here is remarkably robust both in the range of the incident angles and also in frequency. The $T_0$ mode in Fig. (\ref{932kHz}a) fully develops at around 93 kHz and the second mode does not start to develop before 261 kHz. Within this range of 168 kHz there exists the possibility of pure negative refraction in the transmitted signal.
\begin{figure}[htp]
\centering
\includegraphics[scale=.5]{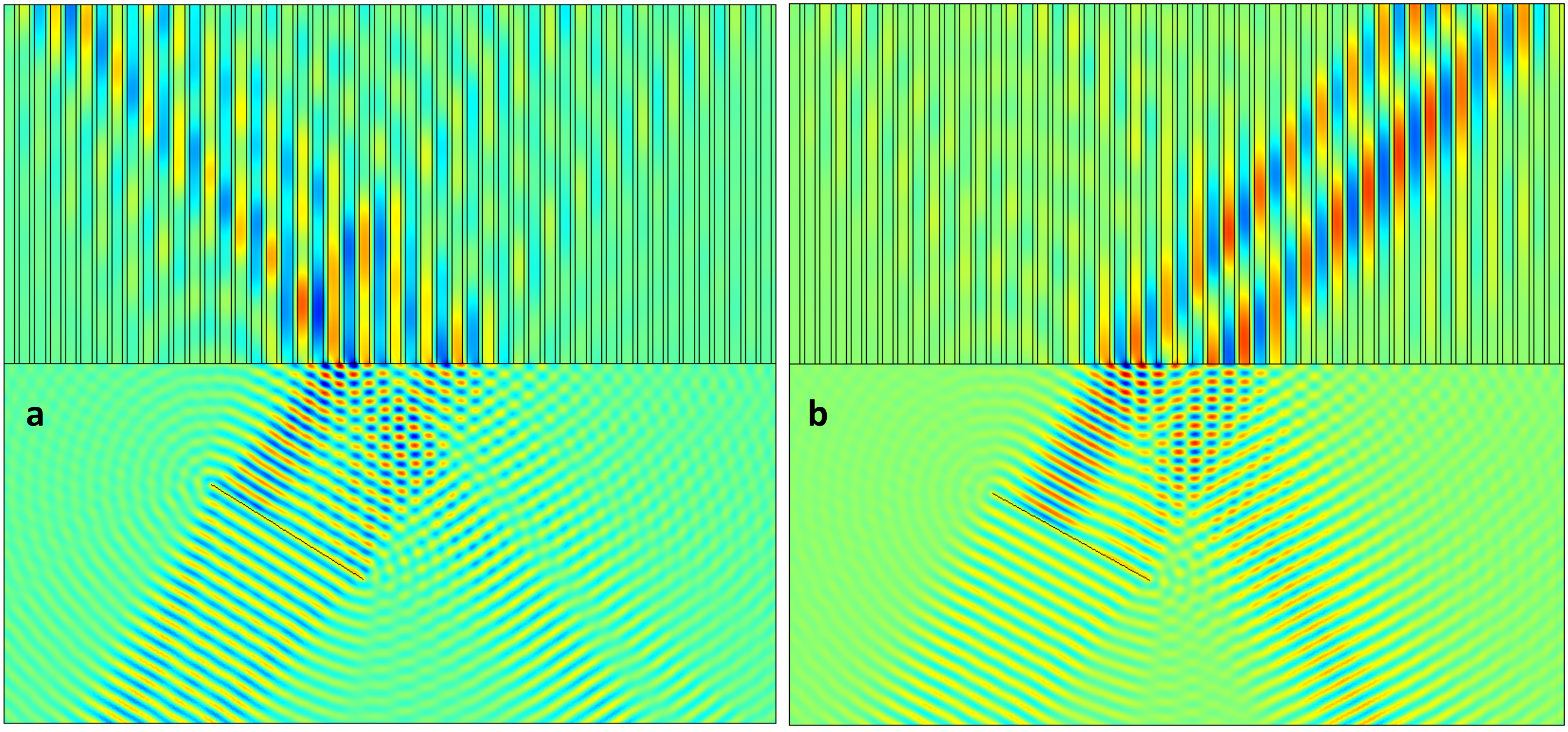}
\caption{Beam steering in the laminate, a. Negative refraction at an angle of incidence of $32^0$, b. Positive refraction at an angle of incidence of $29^0$.}\label{steering}
\end{figure}

Another consequence of the large refraction angle near the high symmetry point for the present example is that the laminate can be used to steer beams with minimal change in the angle of incidence. Fig. (\ref{steering}) shows this phenomenon at the frequency of 93.2 kHz. The incident wave in Fig. (\ref{steering}a) makes an angle of 32$^0$ with the normal. At this incidence angle $K_1h$ is just larger than $\pi$ and, therefore, there is a pure negative refraction in the laminate. The incidence angle of the wave in Fig. (\ref{steering}b) is 29$^0$. This translates into a $K_1h$ value which is slightly less than $\pi$. The transmitted wave, therefore, refracts positively in the laminate and it is the only transmission possible. The transmitted beam traverses roughly an angle of 70$^0$ for only a 3$^0$ change in the angle of the incident beam. 

There are other interesting aspects of this simple problem but we will not present explicit calculations for each. As an example, it is possible to switch the transmitted beam (or a portion of it) on and off with a small change in the angle of incidence. This happens when a frequency is chosen where one of the modes is only partially developed. At frequencies less than 93.2 kHz there are no real $K_1h,k_2h$ solutions for some $\alpha<K_1h<\pi$. By tuning the incident beam to have a $kh\sin(\theta)$ around the value $\alpha$, the only transmitted mode can be cleanly switched on and off. This could obviously be done to higher modes as well. There exists a frequency range for each higher mode where it is only partially developed. Within that frequency range, that particular mode can be individually switched on and off by a small change in the incidence angle.

\section{Laminate as a highpass frequency filter}

When the laminate-homogeneous material interface is along $x_1=0$ then another interesting dispersion characteristic of the laminate can be utilized to practical effect. To understand this effect we first note, from previous theoretical arguments and Fig. (\ref{932kHz}a), that beyond some finite $k_2h$ value there are no real $K_1h$ solutions. Let's call this value $\beta$ which will clearly be dependent upon the frequency under consideration. $\beta$ will increase with frequency but we are only interested in the ratio $\beta/\omega$.
\begin{figure}[htp]
\centering
\includegraphics[scale=.25]{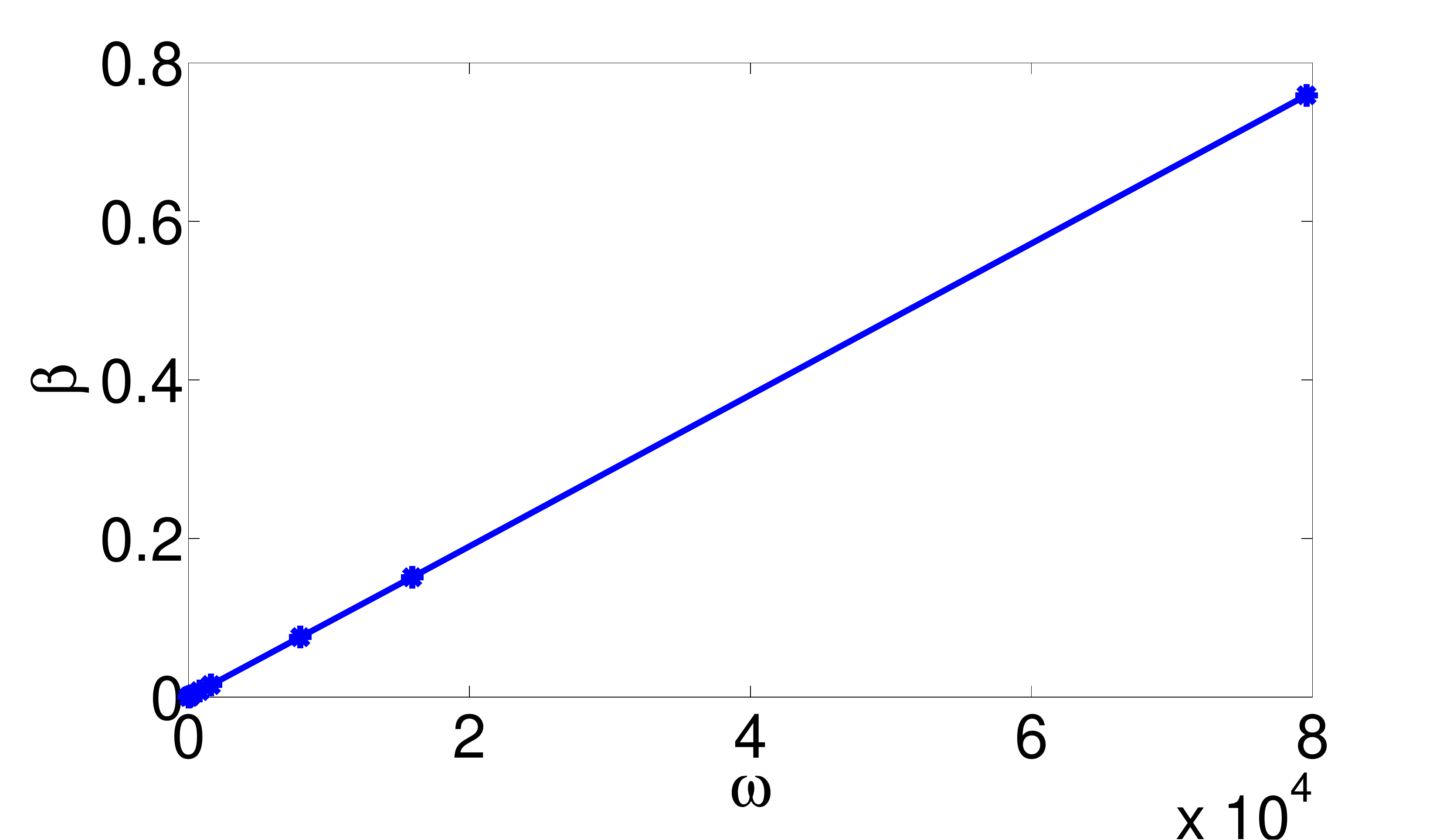}
\caption{$\beta-\omega$ plot for the laminate showing the constancy of $\beta/\omega$ over the chosen frequency range.}\label{bo}
\end{figure}

Fig. (\ref{bo}) shows the variation of $\beta$ with $\omega$ up to an $\omega$ value of $80,000$ rad. It is clear that in this range $\beta/\omega$ is a constant. This ratio starts to depend upon frequency for much higher frequency values, but from static to some maximum value of frequency this ratio may be taken as a constant. Now let's consider the interface problem between the layered composite and a homogeneous medium where the interface is along $x_1=0$. A plane wave with wavenumber $k$ and frequency $\omega$ is incident from the homogeneous medium at the interface and it makes an angle $\theta$ with the normal (with the $x_1$ axis). Clearly, if $kh\sin(\theta)$ is greater than $\beta$ then no propagating wave will be transmitted inside the laminate. This translates into the condition:
\begin{equation}
\frac{1}{c}>\frac{\beta}{h\omega\sin\theta}
\end{equation}
where $c$ is the speed of the wave in the homogeneous medium. 

\begin{figure}[htp]
\centering
\includegraphics[scale=.65]{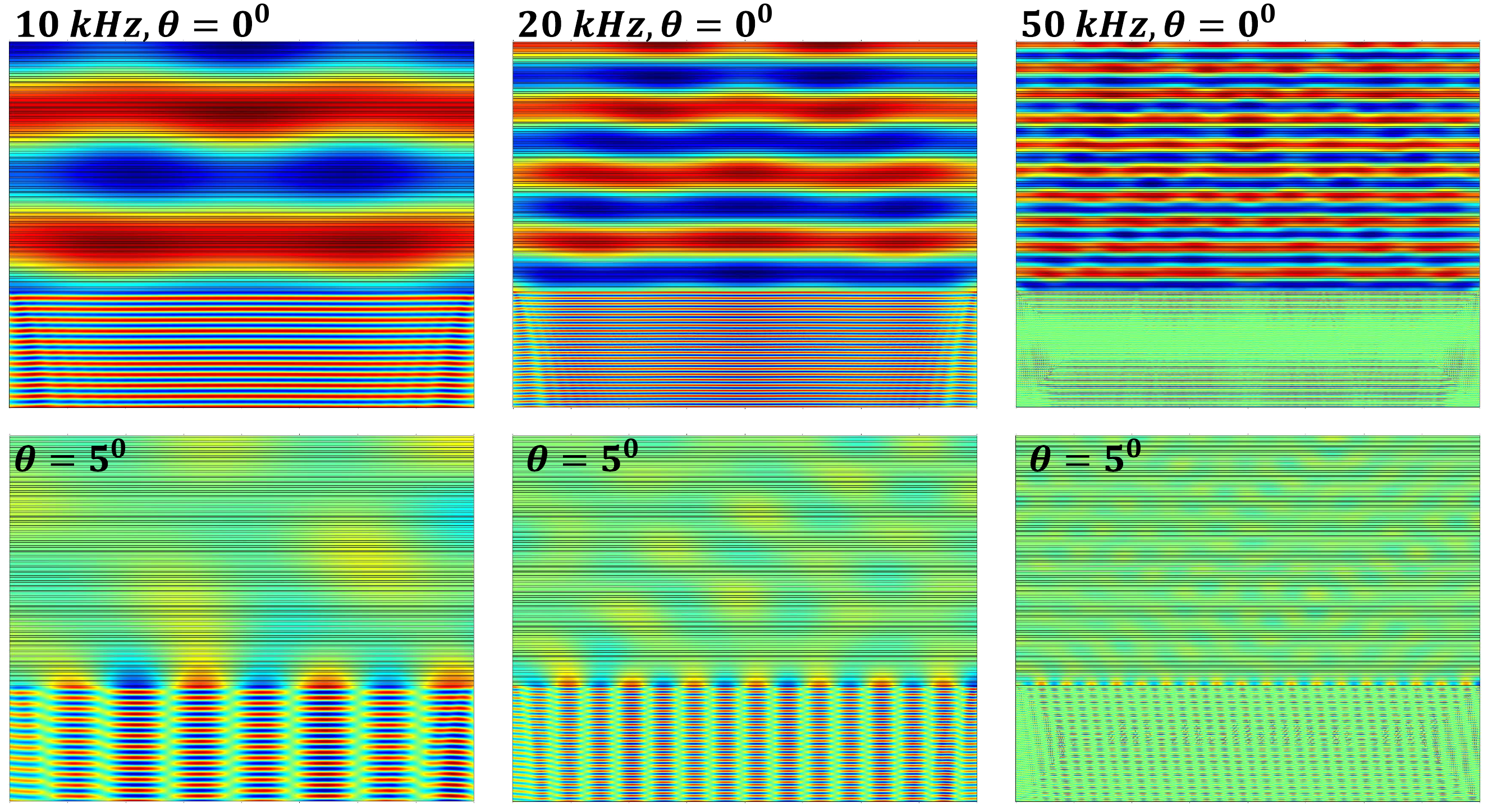}
\caption{Comsol simulation showing the effectiveness of the laminate as a highpass frequency filter. For the three frequencies considered, waves are transmitted for normal incidence but blocked for $5^0$ incident angle.}\label{filter1}
\end{figure}

For a given $\theta$ and in the frequency range where $\beta/\omega$ is a constant this equation will automatically be satisfied everywhere if it is made to satisfy at one frequency. Moreover, if this equation is satisfied for some $\theta_0$ then it will automatically be satisfied for all $\theta>\theta_0$ in this frequency range. A laminate can, therefore, serve to block all plane waves which have a frequency in this range and which impinge the laminate at an angle greater than $\theta_0$.

As a demonstration of the principle the above means that we can choose an arbitrarily small (but non-zero) $\theta_0$ and the equation $1/c=\beta/(h\omega\sin\theta_0)$ will determine the velocity of the wave in the homogeneous medium. Any plane wave traveling in this homogeneous medium which impinges the interface at an angle greater than $\theta_0$ will not be transmitted in the laminate. Moreover, this effect will be a broadband one. It can be made to exist from quasi-static values up to very large values of frequency. It is also interesting to note here that the above argument only fixes the speed of the wave inside the homogeneous medium. The homogeneous medium may very well be impedance matched with the laminate (through homogenization techniques). Even in this extreme case, though, the laminate will reflect most of the waves. This extreme case is demonstrated in Fig. (\ref{filter1}). For the current laminate a rough homogenization can be done through simple volume averaging of density and compliance. This results in an effective impedance value of $I=3.7\times 10^6$ Pa-s/m. We assume the homogeneous material has this impedance. Due to the impedance match, the waves traveling in this homogeneous material will transmit to the layered composite with little reflections, especially for low frequencies and close to normal angles of incidence. Additionally we fix $\theta_0=2^0$. These two requirements uniquely determine the material properties of the homogeneous material ($\rho_0=39488$ kg/m$^3$, $c=93.7924$ m/s). These are unrealistic material properties but they serve to elucidate the concept. Fig. (\ref{filter1}) shows the frequency domain Comsol simulations of the problem under consideration. For the three frequencies considered, it can be seen that the wave easily passes through the interface for normal incidence. This is due to the homoegeneous material being impedance matched with the laminate. Despite the impedance matching, however, for a small non-zero angle of incidence ($\theta=5^0>\theta_0$) the transmission is vastly reduced in each case. This effect becomes more pronounced for larger angles of incidence and persists over larger frequencies (results not shown here). The effect is expected to persist for smaller frequencies as well but accurate computational verification for those cases require larger computational domains - not attempted here.

\section{Conclusions}

Some new observations regarding waves in laminate have been presented here. The observations pertain to an interface between the laminate and a homogeneous material and, as such, can be separated into two categories: a. when the interface is parallel to the layers and, b. when the interface is perpendicular to the layers. For case (a) we have shown that for a given homogeneous medium it may be possible to design a laminate which will act as a high-pass frequency filter for all waves except those which are incident at the laminate within an arbitrarily small range of incident angles close to and including normal incidence. This observation is different from the omnidirectional reflector observation which has already been made in literature. The latter is a band-pass filter whereas ours is a high-pass filter. For case (b) we have shown that it is possible to have a purely negatively refracted signal in the laminate. The effect is persistent over large angles of incidences and frequency ranges. As a corollary we have shown that it is possible to steer beams in the laminate over large angles with small changes in the angle of incidence. Furthermore, it is also possible to switch the beams on and off for small changes in incident angles. In all of this, since it is possible to generate just one propagating wave in a laminate the effects can be controlled with much more ease than in 2-D and 3-D phononic crystals.

\section{Acknowledgments}
The author acknowledges the support of the NSF CAREER grant $\#$1554033 to the Illinois Institute of Technology and UCSD subaward UCSD/ONR W91CRB-10-1-0006 to the Illinois Institute of Technology (DARPA AFOSR Grant RDECOM W91CRB-10–1-0006 to
the University of California, San Diego).

%\bibliography{../References/ReferencesBib}

\begin{thebibliography}{27}%
\makeatletter
\providecommand \@ifxundefined [1]{%
 \@ifx{#1\undefined}
}%
\providecommand \@ifnum [1]{%
 \ifnum #1\expandafter \@firstoftwo
 \else \expandafter \@secondoftwo
 \fi
}%
\providecommand \@ifx [1]{%
 \ifx #1\expandafter \@firstoftwo
 \else \expandafter \@secondoftwo
 \fi
}%
\providecommand \natexlab [1]{#1}%
\providecommand \enquote  [1]{``#1''}%
\providecommand \bibnamefont  [1]{#1}%
\providecommand \bibfnamefont [1]{#1}%
\providecommand \citenamefont [1]{#1}%
\providecommand \href@noop [0]{\@secondoftwo}%
\providecommand \href [0]{\begingroup \@sanitize@url \@href}%
\providecommand \@href[1]{\@@startlink{#1}\@@href}%
\providecommand \@@href[1]{\endgroup#1\@@endlink}%
\providecommand \@sanitize@url [0]{\catcode `\\12\catcode `\$12\catcode
  `\&12\catcode `\#12\catcode `\^12\catcode `\_12\catcode `\%12\relax}%
\providecommand \@@startlink[1]{}%
\providecommand \@@endlink[0]{}%
\providecommand \url  [0]{\begingroup\@sanitize@url \@url }%
\providecommand \@url [1]{\endgroup\@href {#1}{\urlprefix }}%
\providecommand \urlprefix  [0]{URL }%
\providecommand \Eprint [0]{\href }%
\providecommand \doibase [0]{http://dx.doi.org/}%
\providecommand \selectlanguage [0]{\@gobble}%
\providecommand \bibinfo  [0]{\@secondoftwo}%
\providecommand \bibfield  [0]{\@secondoftwo}%
\providecommand \translation [1]{[#1]}%
\providecommand \BibitemOpen [0]{}%
\providecommand \bibitemStop [0]{}%
\providecommand \bibitemNoStop [0]{.\EOS\space}%
\providecommand \EOS [0]{\spacefactor3000\relax}%
\providecommand \BibitemShut  [1]{\csname bibitem#1\endcsname}%
\let\auto@bib@innerbib\@empty
%</preamble>
\bibitem [{\citenamefont {Pennec}\ \emph {et~al.}(2010)\citenamefont {Pennec},
  \citenamefont {Vasseur}, \citenamefont {Djafari-Rouhani}, \citenamefont
  {Dobrzy{\'n}ski},\ and\ \citenamefont {Deymier}}]{pennec2010two}%
  \BibitemOpen
  \bibfield  {author} {\bibinfo {author} {\bibfnamefont {Y.}~\bibnamefont
  {Pennec}}, \bibinfo {author} {\bibfnamefont {J.~O.}\ \bibnamefont {Vasseur}},
  \bibinfo {author} {\bibfnamefont {B.}~\bibnamefont {Djafari-Rouhani}},
  \bibinfo {author} {\bibfnamefont {L.}~\bibnamefont {Dobrzy{\'n}ski}}, \ and\
  \bibinfo {author} {\bibfnamefont {P.~A.}\ \bibnamefont {Deymier}},\
  }\href@noop {} {\bibfield  {journal} {\bibinfo  {journal} {Surface Science
  Reports}\ }\textbf {\bibinfo {volume} {65}},\ \bibinfo {pages} {229}
  (\bibinfo {year} {2010})}\BibitemShut {NoStop}%
\bibitem [{\citenamefont {Lee}\ \emph {et~al.}(2012)\citenamefont {Lee},
  \citenamefont {Singer},\ and\ \citenamefont {Thomas}}]{lee2012micro}%
  \BibitemOpen
  \bibfield  {author} {\bibinfo {author} {\bibfnamefont {J.-H.}\ \bibnamefont
  {Lee}}, \bibinfo {author} {\bibfnamefont {J.~P.}\ \bibnamefont {Singer}}, \
  and\ \bibinfo {author} {\bibfnamefont {E.~L.}\ \bibnamefont {Thomas}},\
  }\href@noop {} {\bibfield  {journal} {\bibinfo  {journal} {Advanced
  materials}\ }\textbf {\bibinfo {volume} {24}},\ \bibinfo {pages} {4782}
  (\bibinfo {year} {2012})}\BibitemShut {NoStop}%
\bibitem [{\citenamefont {Hussein}\ \emph {et~al.}(2014)\citenamefont
  {Hussein}, \citenamefont {Leamy},\ and\ \citenamefont
  {Ruzzene}}]{hussein2014dynamics}%
  \BibitemOpen
  \bibfield  {author} {\bibinfo {author} {\bibfnamefont {M.~I.}\ \bibnamefont
  {Hussein}}, \bibinfo {author} {\bibfnamefont {M.~J.}\ \bibnamefont {Leamy}},
  \ and\ \bibinfo {author} {\bibfnamefont {M.}~\bibnamefont {Ruzzene}},\
  }\href@noop {} {\bibfield  {journal} {\bibinfo  {journal} {Applied Mechanics
  Reviews}\ }\textbf {\bibinfo {volume} {66}},\ \bibinfo {pages} {040802}
  (\bibinfo {year} {2014})}\BibitemShut {NoStop}%
\bibitem [{\citenamefont {Srivastava}(2015)}]{srivastava2015elastic}%
  \BibitemOpen
  \bibfield  {author} {\bibinfo {author} {\bibfnamefont {A.}~\bibnamefont
  {Srivastava}},\ }\href@noop {} {\bibfield  {journal} {\bibinfo  {journal}
  {International Journal of Smart and Nano Materials}\ }\textbf {\bibinfo
  {volume} {6}},\ \bibinfo {pages} {41} (\bibinfo {year} {2015})}\BibitemShut
  {NoStop}%
\bibitem [{\citenamefont {Lin}\ and\ \citenamefont
  {McDaniel}(1969)}]{lin1969dynamics}%
  \BibitemOpen
  \bibfield  {author} {\bibinfo {author} {\bibfnamefont {Y.-K.}\ \bibnamefont
  {Lin}}\ and\ \bibinfo {author} {\bibfnamefont {T.}~\bibnamefont {McDaniel}},\
  }\href@noop {} {\bibfield  {journal} {\bibinfo  {journal} {Journal of
  Engineering for Industry}\ }\textbf {\bibinfo {volume} {91}},\ \bibinfo
  {pages} {1133} (\bibinfo {year} {1969})}\BibitemShut {NoStop}%
\bibitem [{\citenamefont {Faulkner}\ and\ \citenamefont
  {Hong}(1985)}]{faulkner1985free}%
  \BibitemOpen
  \bibfield  {author} {\bibinfo {author} {\bibfnamefont {M.}~\bibnamefont
  {Faulkner}}\ and\ \bibinfo {author} {\bibfnamefont {D.}~\bibnamefont
  {Hong}},\ }\href@noop {} {\bibfield  {journal} {\bibinfo  {journal} {Journal
  of Sound and Vibration}\ }\textbf {\bibinfo {volume} {99}},\ \bibinfo {pages}
  {29} (\bibinfo {year} {1985})}\BibitemShut {NoStop}%
\bibitem [{\citenamefont {Kutsenko}\ \emph {et~al.}(2015)\citenamefont
  {Kutsenko}, \citenamefont {Shuvalov}, \citenamefont {Poncelet},\ and\
  \citenamefont {Darinskii}}]{kutsenko2015tunable}%
  \BibitemOpen
  \bibfield  {author} {\bibinfo {author} {\bibfnamefont {A.}~\bibnamefont
  {Kutsenko}}, \bibinfo {author} {\bibfnamefont {A.}~\bibnamefont {Shuvalov}},
  \bibinfo {author} {\bibfnamefont {O.}~\bibnamefont {Poncelet}}, \ and\
  \bibinfo {author} {\bibfnamefont {A.}~\bibnamefont {Darinskii}},\ }\href@noop
  {} {\bibfield  {journal} {\bibinfo  {journal} {The Journal of the Acoustical
  Society of America}\ }\textbf {\bibinfo {volume} {137}},\ \bibinfo {pages}
  {606} (\bibinfo {year} {2015})}\BibitemShut {NoStop}%
\bibitem [{\citenamefont {Lekner}(1994)}]{lekner1994light}%
  \BibitemOpen
  \bibfield  {author} {\bibinfo {author} {\bibfnamefont {J.}~\bibnamefont
  {Lekner}},\ }\href@noop {} {\bibfield  {journal} {\bibinfo  {journal} {JOSA
  A}\ }\textbf {\bibinfo {volume} {11}},\ \bibinfo {pages} {2892} (\bibinfo
  {year} {1994})}\BibitemShut {NoStop}%
\bibitem [{\citenamefont {Willis}(2015)}]{willis2015negative}%
  \BibitemOpen
  \bibfield  {author} {\bibinfo {author} {\bibfnamefont {J.}~\bibnamefont
  {Willis}},\ }\href@noop {} {\bibfield  {journal} {\bibinfo  {journal}
  {Journal of the Mechanics and Physics of Solids}\ } (\bibinfo {year}
  {2015})}\BibitemShut {NoStop}%
\bibitem [{\citenamefont {Sun}\ \emph {et~al.}(1968)\citenamefont {Sun},
  \citenamefont {Achenbach},\ and\ \citenamefont {Herrmann}}]{sun1968time}%
  \BibitemOpen
  \bibfield  {author} {\bibinfo {author} {\bibfnamefont {C.-T.}\ \bibnamefont
  {Sun}}, \bibinfo {author} {\bibfnamefont {J.}~\bibnamefont {Achenbach}}, \
  and\ \bibinfo {author} {\bibfnamefont {G.}~\bibnamefont {Herrmann}},\
  }\href@noop {} {\bibfield  {journal} {\bibinfo  {journal} {Journal of Applied
  Mechanics}\ }\textbf {\bibinfo {volume} {35}},\ \bibinfo {pages} {408}
  (\bibinfo {year} {1968})}\BibitemShut {NoStop}%
\bibitem [{\citenamefont {Nemat-Nasser}(1972)}]{nemat1972harmonic}%
  \BibitemOpen
  \bibfield  {author} {\bibinfo {author} {\bibfnamefont {S.}~\bibnamefont
  {Nemat-Nasser}},\ }\href@noop {} {\bibfield  {journal} {\bibinfo  {journal}
  {Journal of Applied Mechanics}\ }\textbf {\bibinfo {volume} {39}},\ \bibinfo
  {pages} {850} (\bibinfo {year} {1972})}\BibitemShut {NoStop}%
\bibitem [{\citenamefont {Nemat-Nasser}\ \emph {et~al.}(1975)\citenamefont
  {Nemat-Nasser}, \citenamefont {Fu},\ and\ \citenamefont
  {Minagawa}}]{nemat1975harmonic}%
  \BibitemOpen
  \bibfield  {author} {\bibinfo {author} {\bibfnamefont {S.}~\bibnamefont
  {Nemat-Nasser}}, \bibinfo {author} {\bibfnamefont {F.}~\bibnamefont {Fu}}, \
  and\ \bibinfo {author} {\bibfnamefont {S.}~\bibnamefont {Minagawa}},\
  }\href@noop {} {\bibfield  {journal} {\bibinfo  {journal} {International
  Journal of Solids and Structures}\ }\textbf {\bibinfo {volume} {11}},\
  \bibinfo {pages} {617} (\bibinfo {year} {1975})}\BibitemShut {NoStop}%
\bibitem [{\citenamefont {Hegemier}\ and\ \citenamefont
  {Nayfeh}(1973)}]{hegemier1973continuum1}%
  \BibitemOpen
  \bibfield  {author} {\bibinfo {author} {\bibfnamefont {G.}~\bibnamefont
  {Hegemier}}\ and\ \bibinfo {author} {\bibfnamefont {A.~H.}\ \bibnamefont
  {Nayfeh}},\ }\href@noop {} {\bibfield  {journal} {\bibinfo  {journal}
  {Journal of Applied Mechanics}\ }\textbf {\bibinfo {volume} {40}},\ \bibinfo
  {pages} {503} (\bibinfo {year} {1973})}\BibitemShut {NoStop}%
\bibitem [{\citenamefont {Hegemier}\ and\ \citenamefont
  {Bache}(1973)}]{hegemier1973continuum2}%
  \BibitemOpen
  \bibfield  {author} {\bibinfo {author} {\bibfnamefont {G.}~\bibnamefont
  {Hegemier}}\ and\ \bibinfo {author} {\bibfnamefont {T.}~\bibnamefont
  {Bache}},\ }\href@noop {} {\bibfield  {journal} {\bibinfo  {journal} {Journal
  of Elasticity}\ }\textbf {\bibinfo {volume} {3}},\ \bibinfo {pages} {125}
  (\bibinfo {year} {1973})}\BibitemShut {NoStop}%
\bibitem [{\citenamefont {Rayleigh}(1887)}]{rayleigh1887xvii}%
  \BibitemOpen
  \bibfield  {author} {\bibinfo {author} {\bibfnamefont {L.}~\bibnamefont
  {Rayleigh}},\ }\href@noop {} {\bibfield  {journal} {\bibinfo  {journal} {The
  London, Edinburgh, and Dublin Philosophical Magazine and Journal of Science}\
  }\textbf {\bibinfo {volume} {24}},\ \bibinfo {pages} {145} (\bibinfo {year}
  {1887})}\BibitemShut {NoStop}%
\bibitem [{\citenamefont {Rayleigh}(1917)}]{rayleigh1917reflection}%
  \BibitemOpen
  \bibfield  {author} {\bibinfo {author} {\bibfnamefont {L.}~\bibnamefont
  {Rayleigh}},\ }\href@noop {} {\bibfield  {journal} {\bibinfo  {journal}
  {Proceedings of the Royal Society of London. Series A, Containing Papers of a
  Mathematical and Physical Character}\ }\textbf {\bibinfo {volume} {93}},\
  \bibinfo {pages} {565} (\bibinfo {year} {1917})}\BibitemShut {NoStop}%
\bibitem [{\citenamefont {Joannopoulos}\ \emph {et~al.}(2011)\citenamefont
  {Joannopoulos}, \citenamefont {Johnson}, \citenamefont {Winn},\ and\
  \citenamefont {Meade}}]{joannopoulos2011photonic}%
  \BibitemOpen
  \bibfield  {author} {\bibinfo {author} {\bibfnamefont {J.~D.}\ \bibnamefont
  {Joannopoulos}}, \bibinfo {author} {\bibfnamefont {S.~G.}\ \bibnamefont
  {Johnson}}, \bibinfo {author} {\bibfnamefont {J.~N.}\ \bibnamefont {Winn}}, \
  and\ \bibinfo {author} {\bibfnamefont {R.~D.}\ \bibnamefont {Meade}},\
  }\href@noop {} {\emph {\bibinfo {title} {Photonic crystals: molding the flow
  of light}}}\ (\bibinfo  {publisher} {Princeton university press},\ \bibinfo
  {year} {2011})\BibitemShut {NoStop}%
\bibitem [{\citenamefont {Boudouti}\ and\ \citenamefont
  {Djafari-Rouhani}(2013)}]{boudouti2013one}%
  \BibitemOpen
  \bibfield  {author} {\bibinfo {author} {\bibfnamefont {E.~H.~E.}\
  \bibnamefont {Boudouti}}\ and\ \bibinfo {author} {\bibfnamefont
  {B.}~\bibnamefont {Djafari-Rouhani}},\ }in\ \href@noop {} {\emph {\bibinfo
  {booktitle} {Acoustic Metamaterials and Phononic Crystals}}}\ (\bibinfo
  {publisher} {Springer},\ \bibinfo {year} {2013})\ pp.\ \bibinfo {pages}
  {45--93}\BibitemShut {NoStop}%
\bibitem [{\citenamefont {Winn}\ \emph {et~al.}(1998)\citenamefont {Winn},
  \citenamefont {Fink}, \citenamefont {Fan},\ and\ \citenamefont
  {Joannopoulos}}]{winn1998omnidirectional}%
  \BibitemOpen
  \bibfield  {author} {\bibinfo {author} {\bibfnamefont {J.~N.}\ \bibnamefont
  {Winn}}, \bibinfo {author} {\bibfnamefont {Y.}~\bibnamefont {Fink}}, \bibinfo
  {author} {\bibfnamefont {S.}~\bibnamefont {Fan}}, \ and\ \bibinfo {author}
  {\bibfnamefont {J.}~\bibnamefont {Joannopoulos}},\ }\href@noop {} {\bibfield
  {journal} {\bibinfo  {journal} {Optics letters}\ }\textbf {\bibinfo {volume}
  {23}},\ \bibinfo {pages} {1573} (\bibinfo {year} {1998})}\BibitemShut
  {NoStop}%
\bibitem [{\citenamefont {Fink}\ \emph {et~al.}(1998)\citenamefont {Fink},
  \citenamefont {Winn}, \citenamefont {Fan}, \citenamefont {Chen},
  \citenamefont {Michel}, \citenamefont {Joannopoulos},\ and\ \citenamefont
  {Thomas}}]{fink1998dielectric}%
  \BibitemOpen
  \bibfield  {author} {\bibinfo {author} {\bibfnamefont {Y.}~\bibnamefont
  {Fink}}, \bibinfo {author} {\bibfnamefont {J.~N.}\ \bibnamefont {Winn}},
  \bibinfo {author} {\bibfnamefont {S.}~\bibnamefont {Fan}}, \bibinfo {author}
  {\bibfnamefont {C.}~\bibnamefont {Chen}}, \bibinfo {author} {\bibfnamefont
  {J.}~\bibnamefont {Michel}}, \bibinfo {author} {\bibfnamefont {J.~D.}\
  \bibnamefont {Joannopoulos}}, \ and\ \bibinfo {author} {\bibfnamefont
  {E.~L.}\ \bibnamefont {Thomas}},\ }\href@noop {} {\bibfield  {journal}
  {\bibinfo  {journal} {Science}\ }\textbf {\bibinfo {volume} {282}},\ \bibinfo
  {pages} {1679} (\bibinfo {year} {1998})}\BibitemShut {NoStop}%
\bibitem [{\citenamefont {Bousfia}\ \emph {et~al.}(2001)\citenamefont
  {Bousfia}, \citenamefont {El~Boudouti}, \citenamefont {Djafari-Rouhani},
  \citenamefont {Bria}, \citenamefont {Nougaoui},\ and\ \citenamefont
  {Velasco}}]{bousfia2001omnidirectional}%
  \BibitemOpen
  \bibfield  {author} {\bibinfo {author} {\bibfnamefont {A.}~\bibnamefont
  {Bousfia}}, \bibinfo {author} {\bibfnamefont {E.}~\bibnamefont
  {El~Boudouti}}, \bibinfo {author} {\bibfnamefont {B.}~\bibnamefont
  {Djafari-Rouhani}}, \bibinfo {author} {\bibfnamefont {D.}~\bibnamefont
  {Bria}}, \bibinfo {author} {\bibfnamefont {A.}~\bibnamefont {Nougaoui}}, \
  and\ \bibinfo {author} {\bibfnamefont {V.}~\bibnamefont {Velasco}},\
  }\href@noop {} {\bibfield  {journal} {\bibinfo  {journal} {Surface science}\
  }\textbf {\bibinfo {volume} {482}},\ \bibinfo {pages} {1175} (\bibinfo {year}
  {2001})}\BibitemShut {NoStop}%
\bibitem [{\citenamefont {Manzanares-Mart{\'\i}nez}\ \emph
  {et~al.}(2004)\citenamefont {Manzanares-Mart{\'\i}nez}, \citenamefont
  {S{\'a}nchez-Dehesa}, \citenamefont {H{\aa}kansson}, \citenamefont
  {Cervera},\ and\ \citenamefont
  {Ramos-Mendieta}}]{manzanares2004experimental}%
  \BibitemOpen
  \bibfield  {author} {\bibinfo {author} {\bibfnamefont {B.}~\bibnamefont
  {Manzanares-Mart{\'\i}nez}}, \bibinfo {author} {\bibfnamefont
  {J.}~\bibnamefont {S{\'a}nchez-Dehesa}}, \bibinfo {author} {\bibfnamefont
  {A.}~\bibnamefont {H{\aa}kansson}}, \bibinfo {author} {\bibfnamefont
  {F.}~\bibnamefont {Cervera}}, \ and\ \bibinfo {author} {\bibfnamefont
  {F.}~\bibnamefont {Ramos-Mendieta}},\ }\href@noop {} {\bibfield  {journal}
  {\bibinfo  {journal} {Applied physics letters}\ }\textbf {\bibinfo {volume}
  {85}},\ \bibinfo {pages} {154} (\bibinfo {year} {2004})}\BibitemShut
  {NoStop}%
\bibitem [{\citenamefont
  {Nemat-Nasser}(2015{\natexlab{a}})}]{nemat2015refraction}%
  \BibitemOpen
  \bibfield  {author} {\bibinfo {author} {\bibfnamefont {S.}~\bibnamefont
  {Nemat-Nasser}},\ }\href@noop {} {\bibfield  {journal} {\bibinfo  {journal}
  {Acta Mechanica Sinica}\ }\textbf {\bibinfo {volume} {31}},\ \bibinfo {pages}
  {481} (\bibinfo {year} {2015}{\natexlab{a}})}\BibitemShut {NoStop}%
\bibitem [{\citenamefont {Nemat-Nasser}(2015{\natexlab{b}})}]{nemat2015anti}%
  \BibitemOpen
  \bibfield  {author} {\bibinfo {author} {\bibfnamefont {S.}~\bibnamefont
  {Nemat-Nasser}},\ }in\ \href@noop {} {\emph {\bibinfo {booktitle} {Proc. R.
  Soc. A}}},\ Vol.\ \bibinfo {volume} {471}\ (\bibinfo {organization} {The
  Royal Society},\ \bibinfo {year} {2015})\ p.\ \bibinfo {pages}
  {20150152}\BibitemShut {NoStop}%
\bibitem [{\citenamefont {Notomi}(2000)}]{notomi2000theory}%
  \BibitemOpen
  \bibfield  {author} {\bibinfo {author} {\bibfnamefont {M.}~\bibnamefont
  {Notomi}},\ }\href@noop {} {\bibfield  {journal} {\bibinfo  {journal}
  {Physical Review B}\ }\textbf {\bibinfo {volume} {62}},\ \bibinfo {pages}
  {10696} (\bibinfo {year} {2000})}\BibitemShut {NoStop}%
\bibitem [{\citenamefont {Nemat-Nasser}\ and\ \citenamefont
  {Srivastava}(2011)}]{nemat2011negative}%
  \BibitemOpen
  \bibfield  {author} {\bibinfo {author} {\bibfnamefont {S.}~\bibnamefont
  {Nemat-Nasser}}\ and\ \bibinfo {author} {\bibfnamefont {A.}~\bibnamefont
  {Srivastava}},\ }\href@noop {} {\bibfield  {journal} {\bibinfo  {journal}
  {AIP Adv}\ }\textbf {\bibinfo {volume} {1}},\ \bibinfo {pages} {041502}
  (\bibinfo {year} {2011})}\BibitemShut {NoStop}%
\bibitem [{\citenamefont {Srivastava}\ and\ \citenamefont
  {Nemat-Nasser}(2014)}]{srivastava2014limit}%
  \BibitemOpen
  \bibfield  {author} {\bibinfo {author} {\bibfnamefont {A.}~\bibnamefont
  {Srivastava}}\ and\ \bibinfo {author} {\bibfnamefont {S.}~\bibnamefont
  {Nemat-Nasser}},\ }\href@noop {} {\bibfield  {journal} {\bibinfo  {journal}
  {Wave Motion}\ } (\bibinfo {year} {2014})}\BibitemShut {NoStop}%
\end{thebibliography}

\end{document}